\documentstyle [aps]{revtex}

\tighten

\begin{document}

%\preprint{DUKE-TH-93-46}

%\draft

\title{\bf Parton Equilibration in Relativistic Heavy Ion Collisions}

\author{T. S. Bir\'o$^{1,2}$, E. van Doorn$^1$, B. M\"uller$^1$, M. H.
Thoma$^{1,2}$, and X. N. Wang$^{1,3}$}

\address{$^1$Department of Physics, Duke University, Durham, NC 27708-0305\\
$^2$Institut f\"ur Theoretische Physik, Universit\"at Giessen,
D-6300 Giessen, Germany\\
$^3$Nuclear Science Division, Lawrence Berkeley Laboratory, Berkeley,
CA 94720}

\date{\today}

\maketitle

\begin{abstract}
We investigate the processes leading to phase-space equilibration of
parton distributions in nuclear interactions at collider energies.  We
derive a set of rate equations describing the chemical equilibration
of gluons and quarks including medium effects on the relevant QCD
transport coefficients, and discuss their consequences for parton
equilibration in heavy ion collisions.
\end{abstract}

\vfill
\eject

\section{Introduction}

The mechanisms that precede the formation of a thermalized, locally
deconfined plasma of quarks and gluons in relativistic nuclear
collisions have recently attracted considerable interest, because it
was noticed that the preequilibrium phase may influence the yield of
certain quark-gluon plasma signals, such as lepton pairs and hadrons
containing heavy quarks.  The space-time evolution of quark and gluon
distributions has been investigated in the framework of the parton
cascade model \cite{KGBM}. This model is based on the concept of the
inside-outside cascade \cite{AKM,RCH,JPB} and evolves parton
distributions by Monte-Carlo simulation of a relativistic transport
equation involving lowest-order perturbative QCD scattering and parton
fragmentation.  Numerical studies \cite{KGK,ACJR} have shown that
phase-space equilibration of partons occurs over a
period of 1-2 fm/$c$, and is initially dominated by gluon-induced
processes.

{}From these investigations, and from more schematic considerations, a
picture involving three distinct stages of parton evolution has
emerged \cite{BM,ES,MW,IKJR,GK}:

(1) Gluons ``thermalize'' very rapidly, reaching approximately isotropic
momentum space distributions after a time of
the order of 0.3 fm/$c$.

(2) Full equilibration of the gluon phase space density takes
considerably longer.

(3) The evolution of quark distributions lags behind that of the
gluons, because the relevant QCD cross sections are suppressed by a
factor 2-3.

Although this picture emerges from the numerical simulations
of the parton cascade model \cite{GK}, the complexity of these calculations
makes it difficult to obtain a lucid understanding of the dependence
of the different time scales on various parameters and model assumptions.
It is the goal of our present investigation to derive this insight.

A second, equally important, motivation for our study was the desire
to obtain a better physical understanding of the infrared cut-offs
required in the parton cascade model.  The two cut-off parameters
employed in ref. [1], the minimal transverse momentum transfer $p_0$
in binary parton interactions and the infrared cutoff $z_{\rm min}$
in parton fragmentation processes, were determined by comparison with
cross-sections and particle multiplicities measured in nucleon-nucleon
interactions at high energies.  These values are assumed to reflect the
transition between the perturbative and nonperturbative regimes of QCD
in the normal vacuum.  On the other hand, it is well known that color
screening provides a {\it natural} cut-off of long-range interactions in a
deconfined QCD plasma \cite{EG}, and no artificial cut-off parameters are
required to obtain finite perturbative cross sections in a dense medium
\cite{x}.

A previous study \cite{BMW} has shown that screening effects may be
sufficiently strong immediately after the primary parton collision
events in nuclear collisions at RHIC energies and beyond to provide
the infrared cut-off required in the treatment of secondary parton
interactions.  Improved understanding \cite{GW} of the suppression of
fragmentation processes in a dense QCD plasma (Landau-Pomeranchuk
effect) has also allowed to regard the infrared
cut-off $z_{\rm min}$ as a medium-dependent effect in relativistic
nuclear collisions.  We show here how these medium effects can be
utilized to obtain a parameter-free set of equations, based on
perturbative QCD in a dense partonic medium, that describes the
evolution of quark and gluon distributions towards equilibrium.
Arbitrary cut-off parameters enter only into the description of the
primary semi-hard parton scattering, where one must continue to rely
on a comparison with nucleon scattering data.  After this short
initial phase, however, the approach towards an equilibrated
quark-gluon plasma will be described
without need for arbitrary parameters.

Our paper is structured as follows:  Chapter II discusses initial
parton production in relativistic nuclear collisions.  The evolution
of these partons into locally isotropic, quasi-thermal momentum
distributions is described in Chapter III.  In Chapter IV we derive a
set of rate equations describing the further evolution and chemical
equilibration of gluon and quark distributions.  The influence of the
dense medium on the relevant QCD cross sections is discussed in
Chapter V, and a closed set of rate equations including these
medium modifications is derived.  We explore the solution of these
equations in Chapter VI.  Some consequences for quark-gluon plasma
signatures are briefly discussed in the Conclusions.
\bigskip

\section{Minijet Production}
\medskip

In high energy nucleon-nucleon collisions the production of minijets
with $p_T$ about a few GeV becomes increasingly important at colliding
energies beyond the CERN ISR energy range \cite{CA}.  One would
certainly expect that there could be a fairly large number of minijets
produced in ultrarelativistic heavy ion collisons.  It was estimated
by Kajantie et al. \cite{KK} that minijets could contribute
up to half of the total transverse energy produced in heavy ion
collisions at RHIC energy.  Unlike the soft processes which dominate
the reaction at lower energies, those minijets carry relatively large
transverse momenta and they quickly become incoherent from the rest of
nuclear matter in the fragmentation region.  If a sufficiently high
density is reached and the equilibration proceeds rapidly, those
minijets would eventually lead to a thermalized and locally
deconfined quark gluon plasma.

To estimate the initial parton density, we calculate the minijet production
using the HIJING Monte Carlo model \cite{XNG}. In this model,
perturbative QCD, implemented along the lines of Pythia \cite{SZ}, is
combined together with low $p_T$ phenomenology to describe multiple
minijet production in each binary nucleon-nucleon collision.  One
important quantity in the model is the inclusive jet cross section.
Given the parton structure function, $f_a(x,p_T^2)$ and the perturbative
parton-parton cross section $d\sigma_{ab}$, the differential jet cross
section in nucleon-nucleon collisions can be calculated as \cite{RDF}
\begin{equation}
{d\sigma_{\hbox{\rm jet}}\over dp^2_Tdy_1dy_2} = K \sum_{a,b}
x_1f_a(x_1,p^2_T) x_2f_b(x_2,p_T^2) {d\sigma_{ab} \over d\hat t},
\label{1}
\end{equation}
where the phenomenological factor $K\approx 2$ accounts for higher order
corrections and $x_1,x_2$ are the Feynman variables denoting the
longitudinal momentum fraction carried by a parton.  Since this
cross section diverges and the perturbative approach fails at small
$p_T$, we introduce an infrared cut-off $p_0$ to calculate the
total inclusive jet cross section $\sigma_{\hbox{\rm jet}}(p_0)$.
Introducing $\sigma_{\hbox{\rm soft}}$ for the soft
interactions below the cut-off, we {\it obtain} the inelastic cross section
for nucleon-nucleon collisions in the eikonal approximation \cite{XN},
\begin{equation}
\sigma_{\hbox{\rm in}} = \int
d^2b\left[ 1-e^{-(\sigma_{\hbox{\rm soft}}+\sigma_{\hbox{\rm jet}})
T_N(b)}\right], \label{2}
\end{equation}
where $T_N(b)$ is the partonic overlap function between two nucleons
at impact parameter $b$.  In order to account for the shadowing of the
parton density within a nucleus, we also introduced effective parton
structure functions $f_{a/A}(x,p_T^2)$ which include medium corrections
\cite{XNG}.

To fix the two correlated parameters $p_0$ and $\sigma_{\hbox{\rm
soft}}$, we have assumed constant values of $p_0$ = 2 GeV/$c$ and
$\sigma_{\hbox{\rm soft}}$ = 57 mb.  The resultant total, elastic
and inelastic cross sections agree very well with the experiments from
ISR to Tevatron and cosmic-ray energies \cite{XNMG}.  We want to
emphasize here that the infrared cut-off $p_0$ in the primary
parton-parton scatterings is a phenomenological parameter and is
constrained by the experimental values of the total nucleon-nucleon
cross section.  (As we will discuss below, this infrared singularity of
parton cross sections in a dense partonic medium can be
regulated naturally by the color screening mass.)

Due to the high gluon density at small $x$ and the large gluon-gluon
cross section, the initially produced partons are predominantly
gluons.  It is very important for the following discussion of the
emergence of momentum isotropy that the $p_T$ distributions of the
initially produced gluons are almost exponential.  The results of the
HIJING calculation at RHIC and LHC energies are given in ref. \cite{BMW},
where the rapidity width of the central plateau $Y$, the total number
of the produced gluons $N_G$, and their average transverse momentum
$\langle k_T\rangle$ are listed.  Before we estimate the initial parton
number and energy densities, we next have to address the problem of
thermalization.
\bigskip

\section{Thermalization}
\medskip

We can apply rate equations and particle distribution functions to
describe the system only, when approximate local isotropy in momentum
space is achieved.  At the onset of expansion, scattered partons with
very different rapidities are confined to a highly compressed slab.
The width $\Delta \approx$ $2/p_0$ = 0.2 fm of this slab is determined
by the virtuality of the scattering partons, which in turn is connected
with the transverse momentum cut-off $p_0$ for perturbative parton scattering.

At this point we adopt a realistic ansatz for the phase space distribution
of the scattered gluonic partons.  Following ref. \cite{BMW}, we
define the momentum distribution function as:
\begin{equation}
\label{3}
f\left(\mbox{\boldmath $k$}\right)
=
\frac{1}{d_G \left|\mbox{\boldmath $k$}\right|}\;
g\left(\mbox{\boldmath $k$}_T,y\right),
\end{equation}
where $\mbox{\boldmath $k$}_T$ and $y$ are the transverse momentum and the
rapidity of the gluons respectively, and $d_G = 16$ is the color-spin
degeneracy factor for gluons. Parametrizing the HIJING results
$g\left(\mbox{\boldmath $k$}_T,y\right)$ can be represented by
\begin{equation}
\label{4}
g\left(\mbox{\boldmath $k$}_T,y\right)
=
\frac{1}{2 Y}\left[ \Theta\left(y+Y\right) - \Theta\left(y-Y\right)\right]
\;\;\tilde g (k_T),
\end{equation}
where the function $\tilde g(k_T)$ is explicitly given in ref.
\cite{BMW} for Au + Au collisions at RHIC and LHC energies.
For the spatial part of the distribution function,
we assume that the partons form a homogeneous cylinder of length
$\Delta$ and radius $R$.  The phase space distribution function
then factorizes:
\begin{equation}
\label{7}
F\left(\mbox{\boldmath $k,x$}\right)
=
f\left(\mbox{\boldmath $k$}\right)
D\left(\mbox{\boldmath $x$}\right),
\end{equation}
with $D\left(\mbox{\boldmath $x$}\right)$ given by
\begin{equation}
\label{8}
D\left(\mbox{\boldmath $x$}\right)
= {1\over \pi R^2\Delta}\; \Theta \left( {\Delta^2\over 4} - z^2\right)
\;\Theta (R-r).
\end{equation}
The overall normalization
\begin{equation}
\label{9}
N_G = \int d^{3}k \int d^{3}x \;\; F\left(\mbox{\boldmath $k,x$}\right) = \int
d^3k\;f(\mbox{\boldmath $k$})
\end{equation}
yields gluon rapidity densities $N_G/2Y$ = 110 for RHIC and $N_G/2Y$ =
810 for LHC.

By local isotropy in momentum space we mean that a parton located in
the central region of the collision is surrounded by partons with
different but isotropically distributed momenta. This parton does not
interact with all the partons in the plasma, but only with those within
a distance approximately equal to the parton mean free path.  We can
imagine a little box in the central region the size of the mean free
path $\Lambda_f$.  The momentum of each particle in  the box has
components longitudinal and transverse to the beam, $k_L$ and {\boldmath
$k$}$_T$
respectively. We say that local isotropy in momentum space is
established when the variance $\sigma_{L}$ of the longitudinal momentum
distribution of partons in the box equals that of the transverse momentum
distribution, $\sigma_{T}$.  In the case of an exponential distribution
we call the isotropic parton system (approximately) thermalized.  The
variance $\sigma_T$ is defined as
\begin{equation}
\label{10}
\sigma_T^2 = {1\over 2} \left< \mbox{\boldmath $k$}_{T}^2\right>
=
\frac {1}{2N_G}\int d^{3}k \; \mbox{\boldmath $k$}_{T}^2 f\left(
\mbox{\boldmath $k$}\right).
\end{equation}
At RHIC energy we find $\sigma_{T} = 1.07$ GeV/$c$ using the phase space
distribution of ref. \cite{BMW}, the value at LHC energy is 1.76 GeV/$c$.
For the mean free path $\Lambda_f$ of this initial phase
we take \cite{MFP}:
\begin{equation}
\label{11}
\Lambda_f^{-1} = {3\over 2} \alpha_s\sigma_T,
\end{equation}
corresponding to $\Lambda_f = 0.4$ fm (RHIC) and $\Lambda_f$ = 0.25 fm
(LHC), assuming $\alpha_s$ = 0.3.

As the system expands, mainly in the longitudinal direction, partons with
longitudinal momentum will leave the central region and $\sigma_{L}$ will
decrease as a function of time.
Assuming that the system evolves approximately by free streaming, the
longitudinal position of a parton depends on its rapidity
and time
$$z(t) = z(0) + \tanh (y)t.$$
Here $t$ denotes time in the c.m. system and $z(0)$ is the initial
position of the parton along the beam axis $(\vert z(0)\vert \le
{\Delta\over 2})$.  The phase space distribution $F(\mbox{\boldmath $k$},
\mbox{\boldmath $x$})$ depends accordingly on time,
\begin{equation}
F(\mbox{\boldmath $k$}, \mbox{\boldmath $x$};t) = F(\mbox{\boldmath $k$},
\mbox{\boldmath $x$} -\tanh (y)t \hat{\mbox{\boldmath $z$}}\,;0), \label{12}
\end{equation}
where $\hat{\mbox{\boldmath $z$}}$ is a unit vector along the $z$-direction.
Now we can determine $\sigma_L$ and $\sigma_T$ as functions of time:
\begin{equation}
\sigma_L^2(t) =  \frac{\int d^3x \int d^3k
F(\mbox{\boldmath $k$},\mbox{\boldmath $x$};t) k_L^2}{\int d^3x \int d^3k
F(\mbox{\boldmath $k$},\mbox{\boldmath $x$};t)}, \label{13}
\end{equation}
where the $x$ integral is over $V_f$, the volume of a tube of longitudinal
length $\Lambda_f$, centered at $z=0$.  We find that $\sigma_L(t)$ equals
$\sigma_T$ for $t_{\hbox{\rm iso}}$ = 0.31 fm/$c$ (RHIC) and
$t_{\hbox{\rm iso}}$ = 0.23 fm/$c$ (LHC).  For $t\ge t_{\hbox{\rm
iso}}$, there is local isotropy in momentum space and we will describe the
system by thermal rate equations from then on.

\section{Parton Chemistry}

As discussed in the previous section, the kinematic separation of
partons with different rapidity establishes conditions required
for the validity of continuum dynamics after a short time of the order
of 0.3 fm/$c$.  At this time the momentum space distribution of partons
is roughly isotropic locally and approximately exponential. Since we are
here primarily interested in the chemical equilibration of the parton gas,
we shall assume that the parton distributions can be approximated by
thermal phase space distributions with non-equilibrium fugacities $\lambda_i$:
\begin{equation}
f(k;T,\lambda_i) = \sum_{n=1}^{\infty} \lambda_i^n e^{-n\beta u\cdot
k} =
\lambda_i\left( e^{\beta u\cdot k} \pm
\lambda_i\right)^{-1}, \label{16}
\end{equation}
where $\beta$ is the inverse temperature and $u^{\mu}$ is the four-velocity
of the local comoving reference frame.  The expression (\ref{16}) is
known as J\"uttner distribution.  When the parton fugacities $\lambda_i$
are much less than unity as may happen during the early evolution of the
parton system,  we can neglect the quantum corrections in (\ref{16}) and
write the momentum distributions in the factorized form
\begin{equation}
f(k;T,\lambda _i) \approx \lambda_i e^{-\beta u\cdot k}. \label{17}
\end{equation}

However, this introduces errors of the order of 40 percent, when
the parton phase space distributions approach chemical equilibrium.
As an example, we show the dependence of the screening length
$\mu_D^{-1}$ for static color-electric fields related to the square
of the Debye mass \cite{BMW}
\begin{equation}
\mu_D^2 = {6g^2\over \pi^2} \int_0^{\infty} kf(k) dk, \label{18}
\end{equation}
on the gluon fugacity $\lambda_g$ in Figure 1.
The solid line is the exact perturbative result for the J\"uttner
distribution (\ref{16}), the dotted line shows the Boltzmann approximation
(\ref{17}), and the dashed line shows the result obtained for the factorized
Bose distribution
\begin{equation}
\label{19}
f(k;T,\lambda _i)=\lambda _i\left (e^{\beta u\cdot k}\pm 1\right)^{-1}.
\end{equation}
The dimensionless quantity $\mu_D^2/\lambda_gg^2T^2$ drops from unity
at $\lambda_g=1$ to $6/\pi^2$ at \break $\lambda_g=0$. As can be
seen the Boltzmann distribution function (\ref{17}) is a good approximation
to the gluon phase space density at small $\lambda_g \approx 0$ values.
For $\lambda_g \approx 1$ values, however, the approximation (\ref{19})
applies, which we will adopt in most of the following calculations.
Similar, although smaller, violations of factorization occur in the
gluon and quark densities and in the various reaction rates.

In general, chemical rections among partons can be quite complicated
because of the possibility of initial and final-state gluon radiation.
Multiple radiative emission processes have been investigated in the
framework of numerical simulations of parton cascades \cite{KGBM,KGK}.
However, since radiation processes are considerably suppressed in a
dense parton medium due to rescattering \cite{GW}, we shall here consider only
processes where a single additional gluon is radiated, such as $gg\to
ggg$.  In order to permit approach to chemical equilibrium, the reverse
process, i.e. gluon absorption, has to be included as well, which is
easily achieved making  use of detailed balance.  Closer inspection
shows that gluon radiation is dominated by the process $gg\to ggg$,
because radiative processes involving quarks have substantially smaller
cross sections in perturbative QCD, and quarks are considerably less
abundant than gluons in the initial phase of the chemical evolution of
the parton gas.  Here we are interested in understanding the basic
mechanisms underlying the formation of a chemically equilibrated
quark-gluon plasma, and the essential time-scales.  We hence restrict
our considerations to the dominant reaction mechanisms for the
equilibration of each parton flavour.  These are just four processes
\cite{MSM}:
\begin{equation}
gg \leftrightarrow ggg, \quad gg\leftrightarrow
q\overline{q}.\label{20}
\end{equation}
Other scattering processes ensure the maintenance of thermal
equilibrium $(gg\leftrightarrow gg, \; gq \leftrightarrow gq$, etc.) or
yield corrections to the dominant reaction rates
$(gq\leftrightarrow qgg$, etc.).

Restricting to the reactions (\ref{20}) and assuming that elastic parton
scattering is sufficiently rapid to maintain local thermal
equilibrium, the evolution of the parton densities is governed by the
master equations \cite{MSM}:
\begin{eqnarray}
\partial_{\mu}(n_gu^{\mu}) &= &
n_g (R_{2\to 3} - R_{3\to 2}) - (n_g R_{g\to q} - n_q R_{q\to g})
\label{21}\\
\partial_{\mu} (n_qu^{\mu}) &= &\partial_{\mu} (n_{\bar{q}} u^{\mu})
= n_g R_{g\to q} - n_q R_{q\to g}, \label{22}
\end{eqnarray}
where $R_{2\to 3}$ and $R_{3\to 2}$ denote the rates for the process
$gg \to ggg$ and its reverse, and $R_{g\to q}$ and $R_{q\to g}$ those
for the process $gg \to q\bar{q}$ and its reverse, respectively.
The temperature evolves according to the hydrodynamic equation
\begin{equation}
\partial_{\mu} (\epsilon u^{\mu}) + P\;\partial_{\mu} u^{\mu} = 0,
\label{23}
\end{equation}
where viscosity effects have been neglected \cite{Zim}.

The rates depend on the temperature and the fugacities in a rather
complicated way. In order to make analytical progress, we here assume
factorization of the dependence of the densities on the fugacities,
i.e. eq. (\ref{19}), and postpone a full numerical
evaluation of the rate equations. The rates can then be factorized
in the following simple way:
\begin{eqnarray}
n_g \left(R_{2\to 3} - R_{3\to 2}\right)
&= & {1\over 2}\sigma_3\; n_g^2 \left( 1-\frac{n_g}{\tilde n_g}\right),
\label{24}\\
n_g R_{g\to q} - n_q R_{q\to g}
&= & \sigma_2n_g^2 \left( 1 - {n_qn_{\bar q}\tilde n_g^2
\over \tilde n_q \tilde n_{\bar q}
n_g^2}\right) \label{25}
\end{eqnarray}
where $\sigma_3$ and $\sigma_2$ are thermally averaged cross sections:
\begin{equation}
\sigma_3 = \langle\sigma(gg\to ggg)\rangle, \quad \sigma_2 =
\langle \sigma (gg\to q\bar q)\rangle \label{26}
\end{equation}
and $\tilde{n}_i$ denote the densities for $\lambda_i = 1$.
The equations (\ref{21},\ref{22}) can then be rewritten as
\begin{eqnarray}
\partial_{\mu}(\lambda_g\tilde{n}_g u^{\mu}) &\approx &\tilde{n}_g R_3
\lambda_g (1-\lambda_g)-2\tilde{n}_gR_2\lambda_g
\left( 1- {\lambda_q\lambda_{\bar q}\over\lambda_g^2}\right) ,
\label{27} \\
\partial_{\mu} (\lambda_q \tilde{n}_q u^{\mu}) &=
&\partial_{\mu}(\lambda_{\bar q}\tilde{n}_{\bar q} u^{\mu})
\approx  \tilde{n}_g R_2 \lambda_g \left( 1- {\lambda_q\lambda_{\bar q}\over
\lambda_g^2}\right), \label{28}
\end{eqnarray}
where the density weighted reaction rates $R_3$ and $R_2$ are defined as
\begin{equation}
R_3 = \textstyle{{1\over 2}} \sigma_3 n_g, \quad
R_2 = \textstyle{{1\over 2}} \sigma_2 n_g.  \label{29}
\end{equation}

In order to obtain simple solutions to eqs. (\ref{27},\ref{28}) we will
assume that the expansion of the parton fireball is purely longitudinal,
yielding Bjorken's scaling solution \cite{JDB} of the hydrodynamic
equation (\ref{23}):
\begin{equation}
{d\varepsilon\over d\tau} + {\varepsilon+P\over\tau} = 0, \label{30}
\end{equation}
where $\tau$ is the proper time.  This assumption is expected to be
very well satisfied during the early expansion phase of the fireball,
especially at proper time $\tau \ll R_A$, where $R_A$ is the
transverse radius of the fireball.  Further assuming the
ultrarelativistic equation of state
\begin{equation}
\varepsilon = 3P = \left[a_2\lambda_g + b_2(\lambda_q+\lambda_{\bar
q})\right]\;T^4,
\label{31}
\end{equation}
with $a_2 = 8\pi^2/15$, $b_2 = 7\pi^2N_f/40$, where $N_f$ is the
number of dynamical quark flavors, we obtain from (\ref{30}) a relation
between the proper-time dependence of the temperature and the
fugacities.  Denoting the derivative with respect to $\tau$ by an
overdot, we find
\begin{equation}
{\dot\lambda_g + b(\dot\lambda_q+\dot\lambda_{\bar q})\over
\lambda_g+b(\lambda_q+\lambda_{\bar q})} +
4 {\,\dot T\over T} + {4\over 3\tau} = 0; \label{32}
\end{equation}
where
\begin{equation}
b=b_2/a_2 = {21 N_f\over 64}. \label{33}
\end{equation}
Equation (\ref{32}) is easily integrated, resulting in
\begin{equation}
[\lambda_g + b(\lambda_q+\lambda_{\bar q})]^{3/4} T^3\tau = \hbox{const.}
\label{34}
\end{equation}
For a fully equilibrated quark-gluon plasma $(\lambda_g = \lambda_q =
\lambda_{\bar q}= 1)$ this corresponds to the Bjorken solution $T(\tau) =
T_0(\tau_0/\tau)^{1/3}$.

In the same approximation, i.e. neglecting interactions and using
the factorized distributions, the gluon and quark equilibrium densities
are related to the temperature parameter $T$ by:
\begin{eqnarray}
\tilde{n}_g &= &{16\over\pi^2}\zeta (3) T^3 \equiv a_1T^3,
\label{35} \\
\tilde{n}_q &= &{9 \over 2\pi^2} \zeta (3)  N_f T^3 \equiv b_1T^3.
\label{36}
\end{eqnarray}

We can also rewrite eqs. (\ref{23},\ref{27},\ref{28}) as equations
determining the time dependence of the parameters $T,\lambda_g$ and
$\lambda_q$. In the longitudinal scaling expansion we find:
\begin{equation}
\partial_\mu(n_gu^\mu) = u^\mu\partial_\mu n_g + n_g\partial_\mu u^\mu
= {\partial n_g \over\partial\tau} + {n_g\over\tau}. \label{37}
\end{equation}
After division by $n_g$ and $n_q$, respectively, the rate equations
take the form:
\begin{eqnarray}
{\dot\lambda_g\over\lambda_g} + 3{\dot T\over T} + {1\over\tau} &=
&R_3 (1-\lambda_g)-2R_2 \left(1- {\lambda_q\lambda_{\bar q} \over
\lambda_g^2} \right) \label{38} \\
{\dot\lambda_q\over \lambda_q} + 3{\dot T\over T} + {1\over\tau} &=
&R_2 {a_1\over b_1} \left( {\lambda_g\over \lambda_q} -
{\lambda_{\bar q}\over \lambda_g}\right) \label{39}
\end{eqnarray}
with the additional constraint
\begin{equation}
(\lambda_q-\lambda_{\bar q}) T^3\tau = \hbox{const.} \label{40}
\end{equation}
expressing baryon number conservation.  Being interested in parton
thermalization at very high collision energies, we shall assume baryon
symmetric matter, i.e. $\lambda_q = \lambda_{\bar q}$, which solves
(\ref{40}) trivially.  Equations (\ref{34},\ref{38},\ref{39}) determine the
evolution of $T(\tau), \lambda_g(\tau)$, and $\lambda_q(\tau)$ towards
chemical equilibrium, once the reaction rates $R_2$ and $R_3$ are known.
We now turn to these.
\bigskip

\section{The Equilibration Rates}

The cross sections $\sigma _2$ and $\sigma _3$ contain infrared
singularities if calculated in naive perturbation theory. Recently,
Braaten and Pisarski \cite{BP} proposed an effective perturbation theory,
based on a resummation of subsets of diagrams (hard thermal loops),
which takes screening effects into account and avoids inconsistencies
of the naive perturbation theory at finite temperature. This method results
in using effective propagators and vertices, which show a complicated
momentum dependence.  It has been used, e.g., to calculate the stopping
power \cite{DEDX} and the viscosity \cite{VIS} of the quark-gluon plasma,
which also follow both from elastic parton collisions. In order to calculate
$\sigma _2$ and $\sigma _3$ we will adopt a simplified version of this idea
by introducing momentum independent screening masses into the
propagators, whenever infrared divergences arise otherwise.  In the case
of the stopping power and the viscosity this approximation
provides quantitatively good results.

\subsection{Gluon equilibration}

The rate $R_3$ for the process $gg\to ggg$ depends on the triple
differential radiative gluon-gluon cross section, which can be written
as \cite{GB}
\begin{equation}
{d\sigma^{2\to 3}\over d^2q_{\perp} dy d^2k_{\perp}} \approx
{d\sigma^{2\to 2}\over d^2q_{\perp}} \left[ {C_A\alpha_s\over\pi^2}\;
{q_{\perp}^2\over k_{\perp}^2 (\mbox{\boldmath $k$}_{\perp} -
\mbox{\boldmath $q$}_{\perp})^2}\right],
\label{41}
\end{equation}
where $C_A=3$ is the Casimir operator of the adjoint representation
of SU(3).  Here {\bf k}$_{\perp}$ denotes the transverse momentum and
$y$ the longitudinal rapidity of the radiated gluon, and {\bf q}$_{\perp}$
denotes the momentum transfer in the elastic collision.  The
in-medium cross section for elastic scattering of two gluons \cite{CS}
\begin{equation}
{{d\sigma^{2\to 2}}\over {d^2q_{\perp}}} = C_{gg}
{{2\alpha_s^2}\over{(q_{\perp}^2+\mu_D^2)^2}}, \label{42}
\end{equation}
with $C_{gg} = {9\over 4}$, is screened by the Debye mass.

In the presence of a dense medium the emission of radiation is
suppressed, if the gluons scatter again before the emission is completed
(Landau-Pomeranchuk effect), leading to the condition \cite{MGP}
\begin{equation}
k_{\perp}\Lambda_f > 2\cosh y, \label{43}
\end{equation}
where $\Lambda_f$ is the mean free path of a gluon.  The contribution
from soft radiation is strongly suppressed by this effect.

To obtain the gluon production rate $R_3$ we must integrate the
differential cross section (\ref{41}) over momentum transfer
$\mbox{\boldmath $q$}_{\perp}$ and the phase space of the radiated gluon.
The integrand is singular at $\mbox{\boldmath $k$}_{\perp} =
\mbox{\boldmath $q$}_{\perp}$.  We shall deal with this problem in two
different ways.  First, we shall assume that $\mbox{\boldmath
$k$}_{\perp}\cdot \mbox{\boldmath $q$}_{\perp}$ = 0 in the integrand, i.e.
that the radiation is mostly perpendicular to the reaction plane.
Later we will average over the angle between $\mbox{\boldmath $k$}_{\perp}$
and $\mbox{\boldmath $q$}_{\perp}$, introducing an infrared cut-off to avoid
the singularity.  In the first approach, we replace $(\mbox{\boldmath
$k$}_{\perp} - \mbox{\boldmath $q$}_{\perp})^2$ by $(k_{\perp}^2 +
q_{\perp}^2)$ in the integrand,
effectively suppressing collinear bremsstrahlung.  We then arrive at
the following modified differential cross section:
\begin{equation}
{d\sigma^{2\to 3}\over dq_{\perp}^2 dy dk_{\perp}^2} \approx
{2C_A C_{gg} \alpha_s^3 q_{\perp}^2 \over (q_{\perp}^2 +\mu_D^2)^2
k_{\perp}^2 (k_{\perp}^2 + q_{\perp}^2)}\; \theta (k_{\perp} \Lambda_f -
2\cosh y) \theta(\sqrt{s} - k_{\perp} \cosh y), \label{47}
\end{equation}
where the step functions account for the Landau-Pomeranchuk effect
(\ref{43}) and for energy conservation.  $\langle s\rangle = 18T^2$ is
the average squared center-of-mass energy of two gluons in the thermal gas.
The integrated elastic gluon-gluon cross section in the medium is from
eq. (\ref{42})
\begin{equation}
\sigma^{2\to 2} = 2\pi C_{gg} {\alpha_s^2\over \mu_D^2}, \label{48}
\end{equation}
yielding a fugacity independent mean free path
\begin{equation}
\Lambda_f^{-1} = \sigma^{2\to 2} n_g = {\textstyle{9\over 8}}
a_1 \alpha_sT. \label{49}
\end{equation}
Using these values we evaluate the chemical gluon equilibration rate
$R_3 = {1\over 2} n_g\sigma_3$, as defined in eq. (\ref{28}), numerically.
This rate scales with the temperature linearly but is a complicated
function of the gluon fugacity.  The solid line in Figure 2 shows the
scaled rate $R_3/T$ versus $\lambda_g$ for a coupling constant
$\alpha_s = 0.3$.  The dotted line corresponds to the analytical fit
\begin{equation}
R_3 = 2.1 \alpha_s^2 T \left( 2\lambda_g - \lambda_g^2\right)^{1/2},
\label{50}
\end{equation}
which will be used in solving the time dependent rate equations
discussed in the next section.

We have also studied another solution to the problem posed by the
collinear singularity in eq. (\ref{41}), namely, to add a screening
mass to $(\mbox{\boldmath $k$}_{\perp} - \mbox{\boldmath $q$}_{\perp})^2$.
The integral over the relative angle $\varphi$ between $\mbox{\boldmath
$k$}_{\perp}$ and $\mbox{\boldmath $q$}_{\perp}$ can then be performed
analytically, yielding
\begin{equation}
\frac{1}{2\pi} \int d \varphi [(\mbox{\boldmath $k$}_{\perp} -
\mbox{\boldmath $q$}_{\perp})^2 + \mu^2]^{-1} = [(k_{\perp}^2 + q_{\perp}^2 +
\mu^2) -
4k_{\perp}^2 q_{\perp}^2]^{-1/2}. \label{51}
\end{equation}
The rate $R_3$ obtained in this way for $\mu=\mu_D$ is shown in Figure
2 by the dashed line.  It is somewhat larger than the rate obtained by
setting $\cos \varphi$ = 0, so that (\ref{50}) can be considered as a
conservative estimate.

\subsection{Quark equilibration}

The total cross section for the process $gg\to q\bar q$ is dominated
by the Compton diagrams involving exchange of a virtual light quark in
the $t$- or $u$- channel.  For massless quarks the differential cross
section \cite{CS}
\begin{equation}
{d\sigma_q\over dt} = {\pi\alpha_s^2\over s^2} \left[ {1\over 6} \left(
{u\over t} + {t\over u}\right) - {3\over 8} + {3ut\over 4s^2}\right]
\label{52}
\end{equation}
diverges as $u,t\to 0$, hence the medium-induced effective quark mass
plays a crucial role.  Unfortunately, a complete calculation of the
cross section for the process $gg\to q\bar q$ within the framework of
thermal field theory with hard thermal loop resummation \cite{BP} has not yet
been done.  But since the divergence of the cross section is only
logarithmic in this case, it may be a sufficiently good estimate to
simply substitute the effective thermal quark mass as cutoff in the
divergent integral over momentum transfer.  The thermal quark mass,
using Bose and Fermi equilibrium distributions is \cite{KM},
\begin{equation}
M^2  = \left(\lambda_g + {1\over 2}\lambda_q\right)
{4\pi\over 9} \alpha_s T^2 \label{53}
\end{equation}
yielding the total cross section (for $s\gg M^2$):
\begin{equation}
\sigma_q \approx {\pi\alpha_s^2\over 3s} \left( \ln {s\over M^2} -
{7\over 4}\right)^2. \label{54}
\end{equation}
Integrating over thermal gluon distributions and inserting the average
thermal $\langle s\rangle = 18 T^2$ in the logarithm, we have
\begin{equation}
\sigma_2 \approx N_f\langle\sigma_q\rangle \approx N_f {\pi\alpha_s^2\over
48T^2} \left( \ln {81\over 2\pi\alpha_s\lambda_g} - {7\over 4}\right)^2.
\label{55}
\end{equation}
For $\alpha_s$ = 0.3 in the logarithm and neglecting $\lambda_q$, we
obtain the light quark production rate
\begin{equation}
R_2 = {1\over 2}\sigma_2 n_g \approx 0.064 N_f \;\alpha_s^2 \lambda_g T
\left(\ln {7.5\over\lambda_g}\right)^2. \label{56}
\end{equation}

\section{Results}
\medskip

We have solved the rate equations (\ref{38},\ref{39}) and the
energy conservation equation (\ref{32}) simultaneously by numerical
integration using a fourth order Runge-Kutta method.
The initial conditions for these rate equations are the
number density $n_0 = n(t_{\hbox{\rm iso}})$ and the transverse
energy density $\varepsilon_T=\varepsilon_T (t_{\hbox{\rm iso}})$
of gluonic partons, where
\begin{equation}
n_0 = {1\over \pi R^2 t_{\hbox{\rm iso}}}\; {N_G\over 2Y},
\quad \varepsilon_T = n_0 \langle k_T\rangle. \label{14}
\end{equation}
The total initial gluon energy density is $\varepsilon_0 =
4\varepsilon_T/\pi$.  We obtain $n_0$ = 2.4 fm$^{-3}$ and $\varepsilon_0$
= 3.5 GeV/fm$^3$ at RHIC energy, and $n_0$ = 23 fm$^{-3}$ and
$\varepsilon_0$ = 52 GeV/fm$^3$ at LHC energy.  Quarks contribute a
smaller amount to the initial parton energy density, because the
quark-producing cross sections are smaller in perturbative QCD than
those for gluon production \cite{GLR}.  The total quark
contribution to the energy density we estimate as 30 percent, yielding
total initial energy densities of 5 GeV/fm$^3$ at RHIC and 70
GeV/fm$^3$ at LHC.

The temperature and fugacities at the beginning of hydrodynamical
evolution ($t = t_{\hbox{\rm iso}}$) are obtained assuming a
thermal, but not chemically equilibrated Bose distribution for gluons,
i.e.
\begin{equation}
n_0 = \lambda_g^0 a_1 T_0^3, \quad \varepsilon_0 = \lambda_g^0 a_2
T_0^4 \label{15}
\end{equation}
where $a_1 \approx$ 1.95, $a_2 \approx$ 5.26.  The quark fugacity
is taken as $\lambda_q^0 = \lambda_g^0/5$, corresponding to a ratio
3:1 of the initial gluon number to the combined number of quarks and
antiquarks.  The Table shows these relevant quantities at the moment
$t_{\hbox{\rm iso}}$, for Au + Au collisions at RHIC and LHC energies.

The evolution of temperature and the fugacities are shown Figures 3.
We find that the parton gas cools considerably faster than predicted by
Bjorken's scaling solution $(T^3\tau$ = const.), because the production of
additional partons aproaching the chemical equilibrium state consumes
an appreciable amount of energy.  This can be also deduced from eq.
(\ref{34}): while the fugacities increase, the product $T^3\tau$ must
decrease.

The accelerated cooling on the other hand impedes the chemical
equilibration process, which is more apparent at RHIC (Fig. 3a) than
at LHC energies (Fig. 3b).  In order to see where the perturbative
description of the parton plasma is applicable we investigate the time
evolution of the total energy density, eq. (\ref{29}).  The solid lines
in Fig. 4 correspond to the initial conditions shown in Table I, while the
dashed line corresponds to a more optimistic estimate of the initial
gluon production at RHIC.  One realizes that the perturbative parton plasma
has a lifetime of $1-2$ fm/$c$ at RHIC, while at LHC the plasma may exist
in a deconfined phase as long as $5-6$ fm/$c$. The evolution of the
energy density is remarkably insensitive to the magnitude of the
gluon equilibration rate $R_3$. A change in $R_3$ by a factor 2 is
hardly noticeable in the function $\varepsilon(\tau)$.

The initial conditions obtained in section III depend sensitively on
the low-momentum cut-off $p_0$ used to regularize the mini-jet cross sections.
This parameter has usually been determined from fits to the inelastic
nucleon-nucleon cross section.  Ambiguities of this method allow for a
range of $1.4-2$ GeV for the low-momentum cut-off at RHIC energy.  Using
the smaller value \cite{AGM} one obtains an initial gluon fugacity
$\lambda_g \approx 0.18$, which is three times larger than the value
discussed above.  Using this value we find a larger initial energy density
and hence a longer lifetime of 2.5 fm/$c$ for the plasma at RHIC (see
the dashed curve in Fig. 4).  A higher value of $\lambda_g$ at
$t_{\hbox{\rm iso}}$ can also result from rescattering of gluons
during the ``free-streaming'' period between the end of initial
semihard scattering and $t_{\hbox{\rm iso}}$.  Indeed, the parton
cascade simulations \cite{KGK} indicate a rapid increase of gluon
density during this pre-thermal phase.   We conclude that the validity
of this scenario of quark-gluon plasma production by perturbative parton
scattering is uncertain at RHIC energies, but very likely in the LHC
energy domain.

{}From our investigation emerges the following scenario of a nuclear
collision at collider energies:  Within $0.2-0.3$ fm/$c$ a dense parton
gas is produced at central rapidities, which can be described as a
locally thermalized, but not chemically equilibrated quark-gluon
pasma.  In the fragmentation region, where the parton gas created by
minijets is not sufficiently dense to screen color fields, strings and
color ropes are formed \cite{AGS,BNK}, which decay by non-perturbative
QCD processes on a time scale of about 1 fm/$c$.  These strings extend
between the wounded nuclei and the surface of the parton plasma,
penetrating up to a screening length determined by the actual Debye mass.

As in the Bjorken scenario, the scaling expansion of the system can
best be described by using the proper time $\tau$ and space-time
rapidity
\begin{equation}
\eta = {\textstyle{1\over 2}} \ln {t+x\over t-x}
\end{equation}
variables.  This scenario is schematically shown in Fig. 5, where the
physically different space-time regions of the pre-thermal creation
phase, the parton plasma, and the string-dominated phase are indicated.

The main difference to the conventional Bjorken scenario \cite{JDB} is
the far shorter formation time in the central rapidity region,
corresponding to a much higher initial temperature.  The Bjorken
scenario, however, most probably applies to the fragmentation region,
where the non-perturbative color ropes may also decay into a
quark-gluon plasma.  At the interface between the two regions a
reheating of the parton plasma may occur by dissipation of field energy
stored in the strings.  The dissipation of field energy in a
quark-gluon plasma was recently studied by Eskola and Gyulassy
\cite{EGy}.  At RHIC energies, where the approach discussed here leads
to a rather low initial density of partons, color flux tubes may still
permeate the whole central region.  This would result in multiple
interactions between partons from minijet cascades and the soft
component modeled by color strings or ropes.  In addition, the
temperature falls so rapidly that it is difficult to maintain thermal
equilibrium by elastic two-body collisions among partons. At RHIC
energies the characteristic collision rate, given by the inverse
mean free path $\Lambda_f^{-1}$, is comparable to the cooling rate
$\dot T/T$.  As discussed by the authors of ref. [38],
such a state would exhibit large viscosity effects which can
substantially slow down the cooling rate of the QCD plasma.

\section{Conclusions}
\medskip

Our results support the picture emerging from numerical simulations of
parton cascades and simple estimates that the highly excited parton
plasma created in ultrarelativistic nuclear collisions is initially
mainly a gluon plasma.  Our results have important implications for
several experimental signals associated with quark-gluon plasma formation,
as has been pointed out before \cite{ES,MW,IKJR,GK,GKp,SX}:  Rapid gluon
thermalization at a high initial temperature leads to a substantial
thermal contribution to the total yield of charmed quarks.  This
increase comes on top of a substantial suppression of primary charm
production by nuclear shadowing of gluon distributions in the central
rapidity region.  Hence the unambiguous observation of this charm
enhancement will require the comparative study of $p+p$, $p+A$, and
$A+A$ collisions at the same center-of-mass energy.

The lack of chemical equilibration in the quark and gluon densities,
on the other hand, causes a severe depletion of the number of emitted
lepton-pairs compared with the naive thermal estimate.  However, the
shift towards higher invariant masses of the steeply falling
lepton-pair spectrum due to the larger initial temperature in the plasma
can conceivably offset this suppression \cite{GKp,SX}.

Reliable quantitative predictions for these experimental signatures
require complete microscopic simulations of the parton cascade which
include the medium modification of cross-section and effective masses
discussed in this paper.  As we have shown, these medium effects slow the
chemical equilibration of the parton plasma down.  How can the medium
dependence of the parton cross sections be included in microscopic
simulations?  Since these calculations trace the full space-time
evolution of parton densities, it would be straightforward to calculate
the color screening length, as well as the effective medium-induced
masses of collective gluon and quark modes (plasmons and plasminos) as
function of time.  These quantities could then be used
as input parameters for the determination of medium-dependent
scattering cross sections and branching rates.

An unresolved issue remains in the medium-dependence of the lower
cut-off $p_0$ used to regularize perturbative QCD cross sections, which
was fixed at $p_0$ = 2 GeV/$c$ in HIJING and at a similar, but
energy-dependent value in the parton cascade.  One possible way of
approach would be to identify an upper limit to the (medium-dependent)
running QCD coupling constant, above which perturbative cross sections
cannot be used.  This limit $\alpha_s^{\hbox{\rm max}}$ could be
chosen such that it equals $\alpha_s(p_0)$ in vacuum; but it would
correspond to a different momentum cut-off, or even no cut-off at all,
at high parton densities \cite{kg}.

\begin{acknowledgements}
We would like to thank K. Kinder-Geiger for his careful reading of the
manuscript and several helpful discussions on the parton cascade.  This
work was partially supported by the U.S. Department of Energy (grant
DE-FG05-90ER40592) and the N.C. Supercomputing Center.
\end{acknowledgements}
\vfill
\eject

\begin{center}

\begin{table}

\begin{tabular}{|l|r|r|}
&RHIC &LHC \\ \hline
$t_{\hbox{\rm iso}}$ (fm/$c$) &0.31\phantom{0} &0.23\phantom{0} \\
$\varepsilon_0$ (GeV/fm$^3$) &3.5\phantom{00} &52\phantom{.000} \\
$n_0$ (fm$^{-3}$) &2.4\phantom{00} &23\phantom{.000} \\
$\varepsilon_{\hbox{\rm tot}}$ (GeV/fm$^3$) &5\phantom{.000}
&70\phantom{.000} \\
$\Lambda_f$ (fm) &0.40\phantom{0} &0.24\phantom{0} \\
$\langle k_T\rangle$ (GeV) &1.17\phantom{0} &1.76\phantom{0} \\
$T_0$ (GeV) &0.55\phantom{0} &0.83\phantom{0} \\
$\lambda_g^0$ &0.06\phantom{5} &0.16\phantom{0} \\
$\lambda_q^0$ &0.012 &0.03\phantom{0} \\
\end{tabular}
\bigskip

\caption{Values of the relevant parameters characterizing the parton
plasma at the moment $t_{\hbox{\rm iso}}$, when local isotropy
of the momentum distribution is first reached.}

\end{table}

\end{center}

\hfill
\eject

\vfill

\newpage

\begin{figure}
\caption{The square of the thermal screening mass $\mu_D$ of color
electric fields scaled by the equilibrium value $g^2T^2$ is depicted
as function of the gluon fugacity $\lambda_g$.  The solid curve shows
the correct result for the J\"uttner distribution (\protect\ref{16}), the
dashed line corresponds to the factorized Bose distribution
(\protect\ref{19}), while the dotted line indicates the Boltzmann
approximation (\protect\ref{17}) applying in the limit of a very dilute gluon
plasma.}\label{f1}
\end{figure}

\begin{figure}
\caption{The scaled gluon production rate $R_3/T$ is shown
as function of the gluon fugacity $\lambda_g$ for $\alpha_s$ = 0.3.  The
solid line shows the numerical result, using the approximation (42),
while the dotted line shows the analytical fit (45).  The dashed line
represents the result obtained with the screening approximation (46)
to the collinear singularity in the bremsstrahlung cross section.}\label{f2}
\end{figure}

\begin{figure}
\caption{Time evolution of the temperature $T$ and the
fugacities $\lambda_g$ and $\lambda_q$ of gluons and quarks in the
parton plasma created in Au + Au collisions at (a) the RHIC c.m. energy
of 100 GeV/nucleon, (b) the LHC c.m. energy of 3000 GeV/nucleon.  The initial
values for $T,\; \lambda_g$ and $\lambda_q$ are determined from
simulations with the HIJING code and are listed in the Table.}\label{f3}
\end{figure}

\begin{figure}
\caption{Time dependence of the energy density of the parton
plasma at RHIC and LHC energies.  The solid lines show the results
obtained with the initial values taken from the Table.  The dashed
line shows the effect of a three times higher initial gluon density at
pRHIC energy, corresponding to QCD cut-off $p_0\approx$ 1.5 GeV instead
of $p_0$ = 2 GeV.  Whereas a well-developed parton plasma is predicted
to be formed at the LHC, the prediction for RHIC is uncertain.}\label{f4}
\end{figure}

\begin{figure}
\caption{Schematic space-time diagram of the evolution of QCD
matter in the initial phase of an ultrarelativistic heavy ion
collision.  The central rapidity region is dominated by a thermalized,
but not chemically equilibrated parton plasma after $t_{\hbox{\rm
iso}} \approx 0.2 - 0.3$ fm/$c$.}\label{f5}
\end{figure}


\begin{references}
\bibitem{KGBM}
K. Geiger and B. M\"uller, {\sl Nucl. Phys. {\bf B369}}, 600 (1992);
K. Geiger, {\sl Phys. Rev. {\bf D47}}, 133 (1993);  D. Boal, {\sl Phys.
Rev. {\bf C33}}, 2206 (1986).

\bibitem{AKM}
R. Anishetty, P. K\"ohler, and L. McLerran, {\sl Phys. Rev. {\bf D22}}, 2793
(1980).

\bibitem{RCH}
R. C. Hwa and K. Kajantie, {\sl Phys. Rev. Lett. {\bf 56}}, 696 (1986).

\bibitem{JPB}
J. P. Blaizot and A. H. Mueller, {\sl Nucl. Phys. {\bf B289}}, 847
(1987).

\bibitem{KGK}
K. Geiger, {\sl Phys. Rev. {\bf D46}}, 4965 and 4987 (1992).

\bibitem{ACJR}
I. Kawrakow, H.-J. M\"ohring, and J. Ranft, {\sl Nucl. Phys. {\bf A544}},
471c (1992).

\bibitem{BM}
B. M\"uller, Duke Univ. preprint DUKE-TH-92-36, to appear in: {\sl Particle
Production in Highly Excited Matter}, to be published by Plenum Press,
1993.

\bibitem{ES}
E. Shuryak, {\sl Phys. Rev. Lett. {\bf 68}}, 3270 (1992).

\bibitem{MW}
B. M\"uller and X. N. Wang, {\sl Phys. Rev. Lett. {\bf 68}}, 2437
(1992).

\bibitem{IKJR}
I. Kawrakow and J. Ranft, Univ. Leipzig preprint UL-HEP-92-08;
B. K\"ampfer and O. P. Pavlenko, {\sl Phys. Lett. {\bf B289}}, 127
(1992).

\bibitem{GK} K. Geiger and J. Kapusta, University of Minnesota
preprint (UMSI-Rep. 1/93).

\bibitem{EG}
See e.g.: G. Baym, H. Monien, C. J. Pethick, and D. G. Ravenhall, {\sl
Phys. Rev. Lett. {\bf 64}}, 1867 (1990).

\bibitem{x} With the possible exception of processes sensitive to
static color-magnetic interactions.  Although such interactions are
believed to be screened at momenta below the order of $(g^2T)$, a
consistent analytical theory of color-magnetic screening does not yet
exist.


\bibitem{BMW} T.S. Bir\'o, B. M\"uller, and X.N. Wang, {\sl Phys. Lett.
{\bf B283}}, 171 (1992).


\bibitem{GW} M. Gyulassy and X.N. Wang, manuscript in preparation.

\bibitem{CA} C. Albajar et al. {\sl Nucl. Phys. {\bf B309}}, 405
(1988).

\bibitem{KK} K. Kajantie, P. V. Landshoff, and J. Lindfors, {\sl Phys.
Rev. Lett. {\bf 59}}, 2527 (1987); K. J. Eskola, K. Kajantie, and J.
Lindfors, {\sl Nucl. Phys. {\bf B323?}}, 37 (1989).

\bibitem{XNG} X.-N. Wang and M. Gyulassy, {\sl Phys. Rev. {\bf D44}},
3501 (1991).

\bibitem{SZ} P. Sj\"ostrand and M. van Zijl, {\sl Phys. Rev. {\bf
D36}}, 2019 (1987).

\bibitem{RDF} See e.g. R. D. Field, Applications of Perturbative QCD,
{\sl Frontiers in Physics}, Vol. 77 (Addison-Wesley, Redwood City,
1989).

\bibitem{XN} X.-N. Wang, {\sl Phys. Rev. {\bf D43}}, 104 (1991).

\bibitem{XNMG} X.-N. Wang and M. Gyulassy, {\sl Phys. Rev. {\bf D45}},
844 (1992).

\bibitem{GLR} L. V. Gribov, E. M. Levin, M. G. Ryskin, {\sl Phys. Rep.
{\bf 100}}, 1 (1983).

\bibitem{MFP} For justification of this choice see eq. (\ref{49}).  We have
set $a_1\approx 2$ and $T\approx {2\over 3}\sigma_T$.

\bibitem{MSM} T. Matsui, B. Svetitsky, and L. McLerran, {\sl Phys.
Rev. {\bf D34}}, 783 (1986).

\bibitem{Zim} P. Danielewicz and M. Gyulassy, {\sl Phys. Rev. {\bf D31}},
53 (1985); A. Hosoya and K. Kajantie, {\sl Nucl. Phys. {\bf B250}},
666 (1985); S. Gavin, {\sl Nucl. Phys. {\bf A435}}, 826 (1985).

\bibitem{JDB} J. D. Bjorken, {\sl Phys. Rev. {\bf D27}}, 140 (1983).

\bibitem{BP} E. Braaten and R.D. Pisarski, {\sl Nucl. Phys. {\bf B337}},
569 (1990).

\bibitem{DEDX} M.H. Thoma and M. Gyulassy, {\sl Nucl. Phys. {\bf B351}},
491 (1991); E. Braaten and M.H. Thoma, {\sl Phys. Rev. {\bf D44}}, 1298
and 2625 (1991); M.H. Thoma, {\sl Phys. Lett. {\bf B273}}, 128 (1991).

\bibitem{VIS} M.H. Thoma, {\sl Phys. Lett. {\bf B269}}, 144 (1991).

\bibitem{GB} J.F. Gunion and G. Bertsch, {\sl Phys. Rev. {\bf D25}}, 746
(1982).

\bibitem{CS} R. Cutler and D. Sivers, {\sl Phys. Rev. {\bf D17}}, 196
(1978).

\bibitem{MGP} M. Gyulassy, M. Pl\"umer, M. H. Thoma, and X. N. Wang,
{\sl Nucl. Phys. {\bf A538}}, 37c (1992).  [Note that the factor 2
appearing in eq. (39) is missing in this reference.]

\bibitem{KM} V. V Klimov, {\sl Sov. Phys. JETP {\bf 55}}, 199 (1982);
H.A. Weldon, {\sl Phys. Rev. {\bf D26}}, 2789 (1982).

\bibitem{AGM} N. Abou-El-Naga, K. Geiger, and B. M\"uller,
{\sl J. Phys. {\bf G18}}, 797 (1992).

\bibitem{AGS} See e.g.: B. Andersson, G. Gustafsson, and T.
Sj\"ostrand, {\sl Z. Phys. {\bf C6}}, 235 (1980).

\bibitem{BNK} T. Bir\'o, H. B. Nielsen, and J. Knoll, {\sl Nucl. Phys.
{\bf B245}}, 449 (1984).

\bibitem{EGy} K. J. Eskola and M. Gyulassy, preprint LBL-33150 (December
1992).

\bibitem{GKp} K. Geiger and J. I. Kapusta, University of Minnesota
preprint (January 1993).

\bibitem{SX} E. Shuryak and L. Xiong, Stony Brook preprint
SUNY-NTG-92-37.

\bibitem{kg} This idea was developed in discussions with K. Geiger.

\end{references}
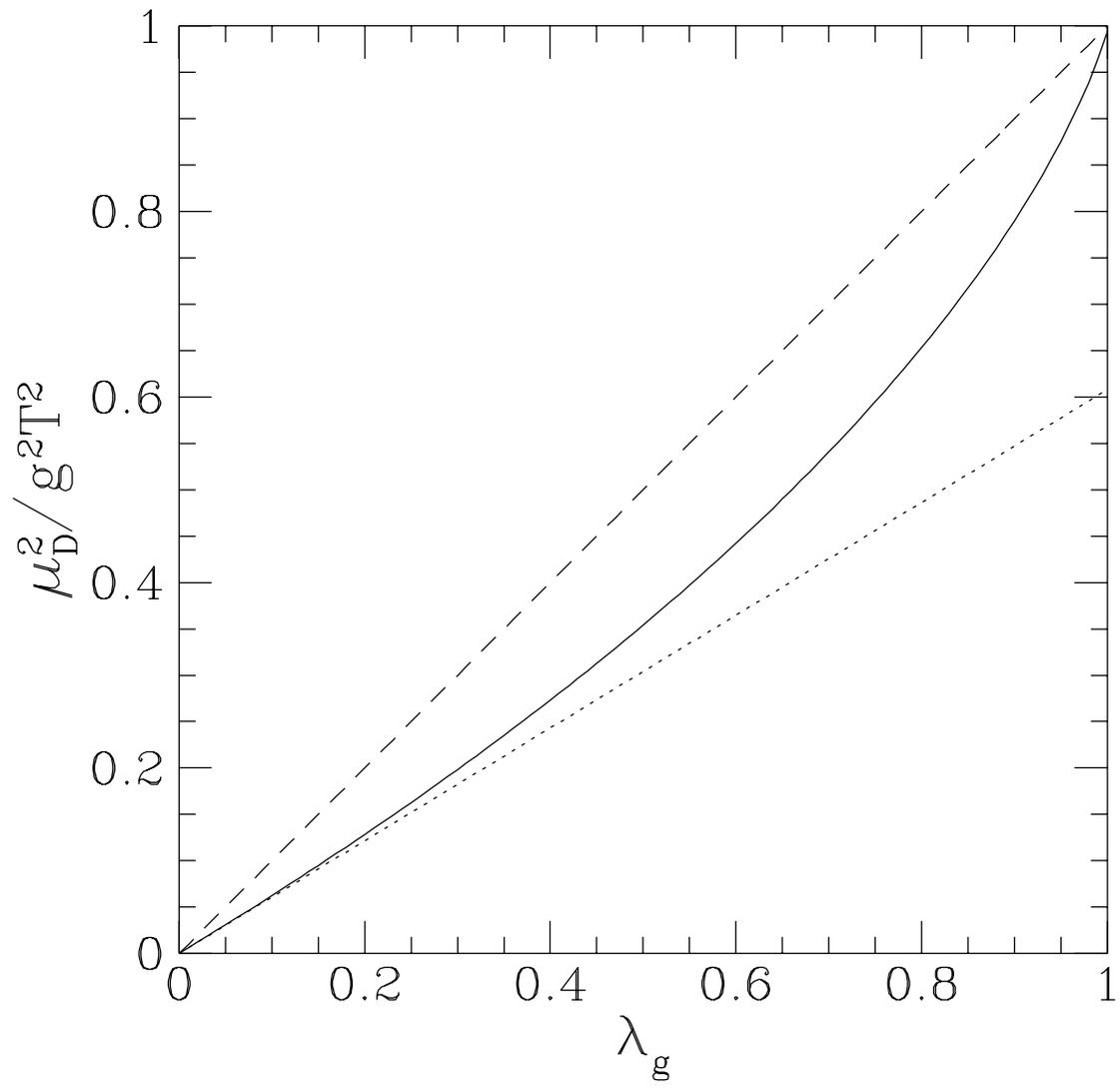
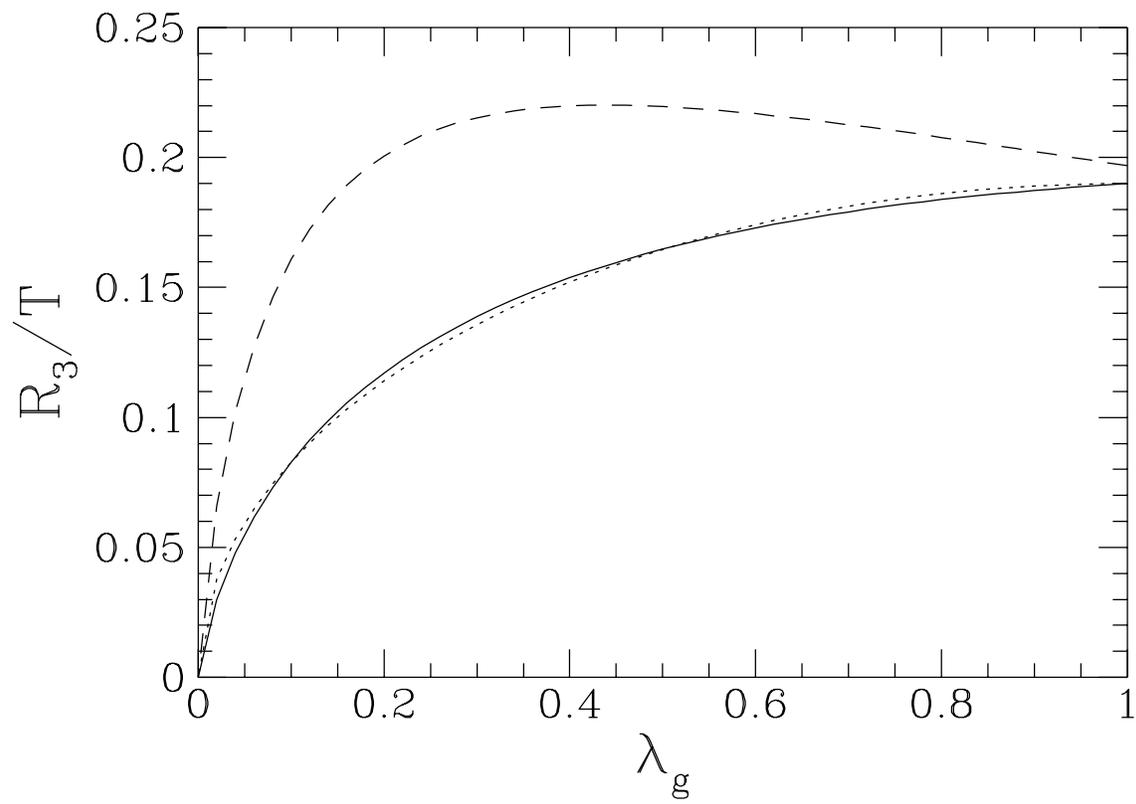
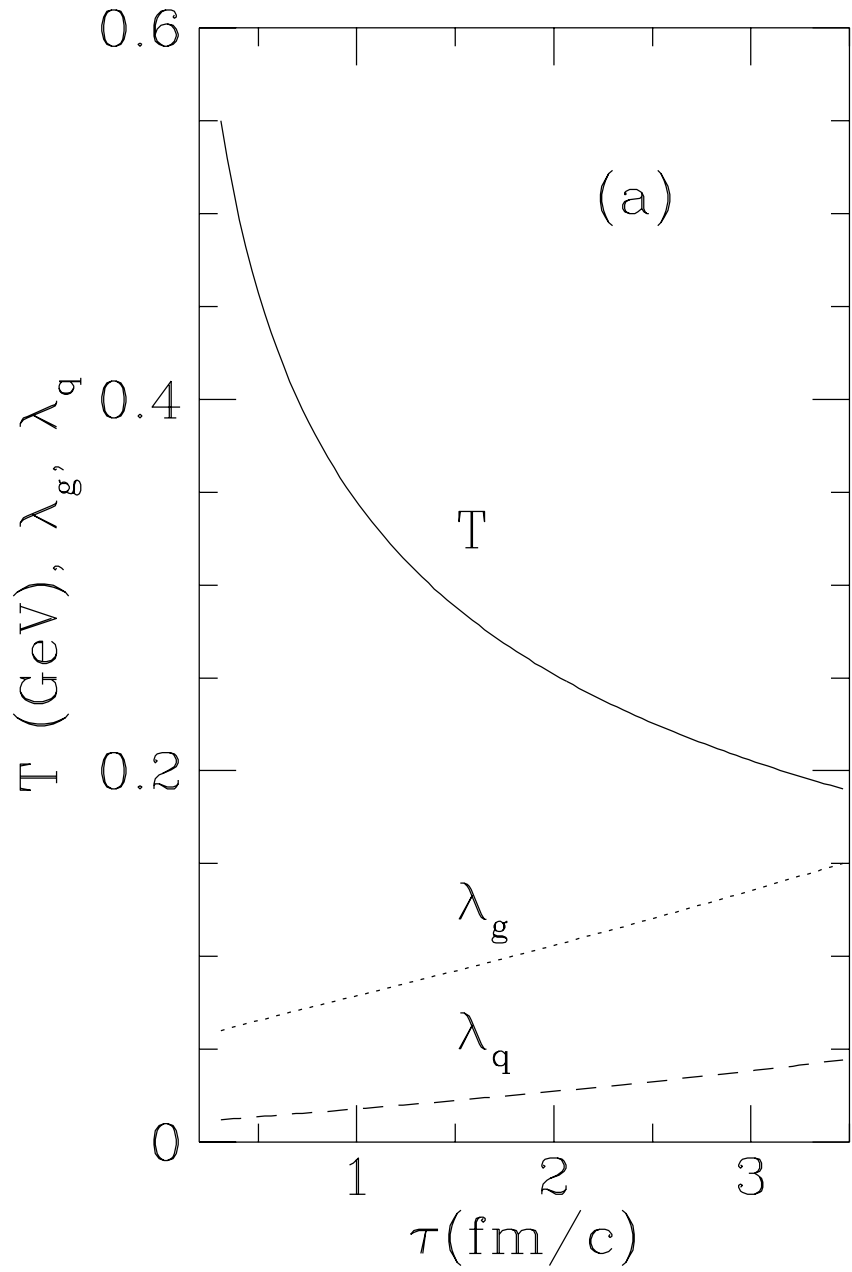
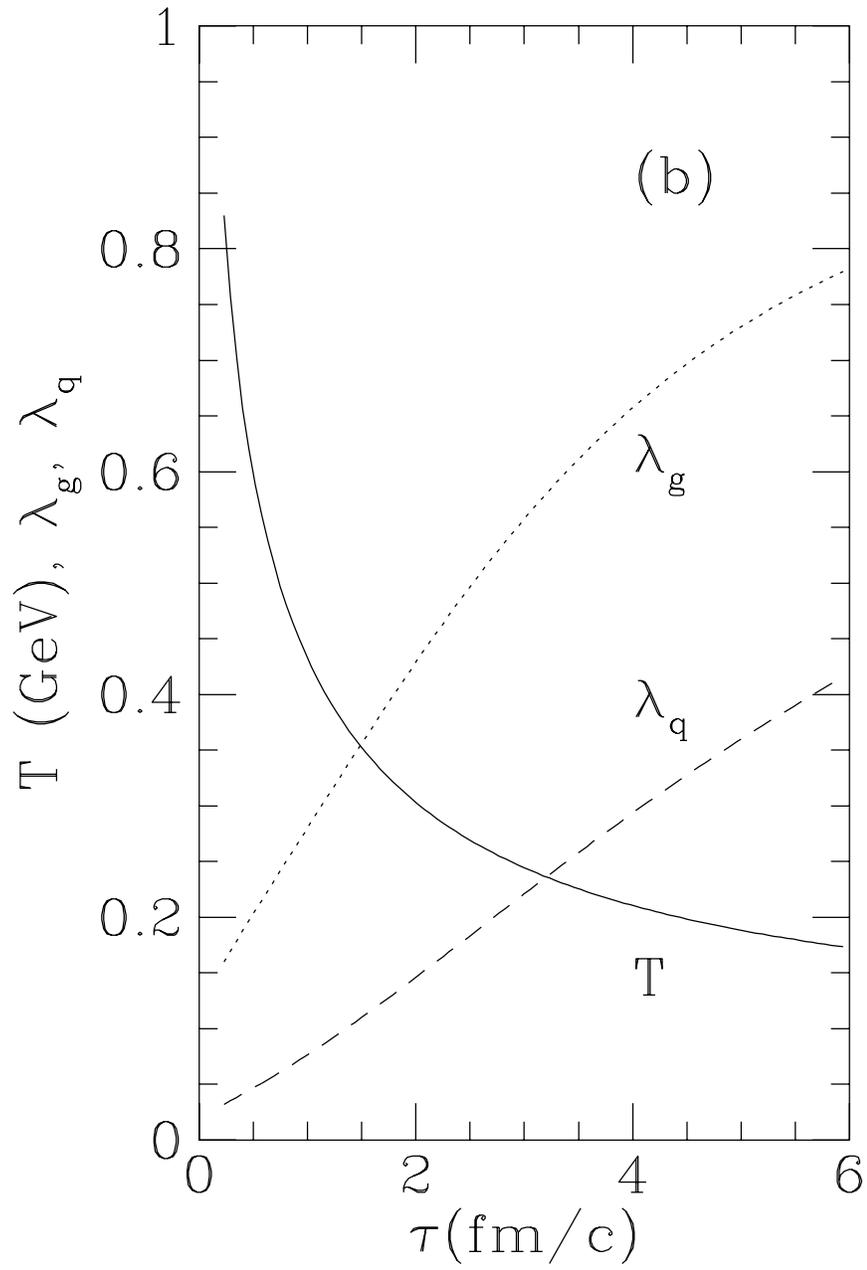
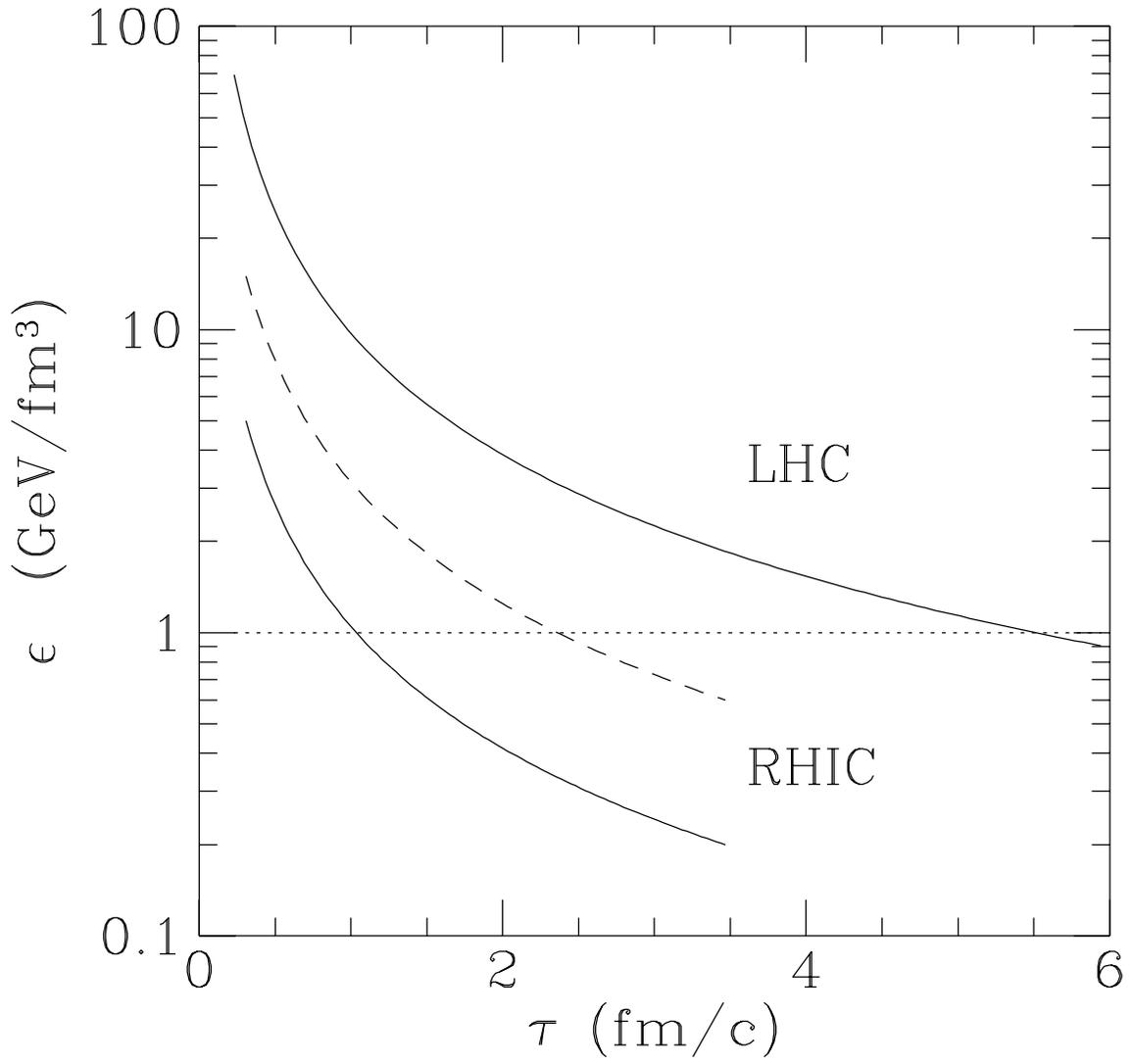
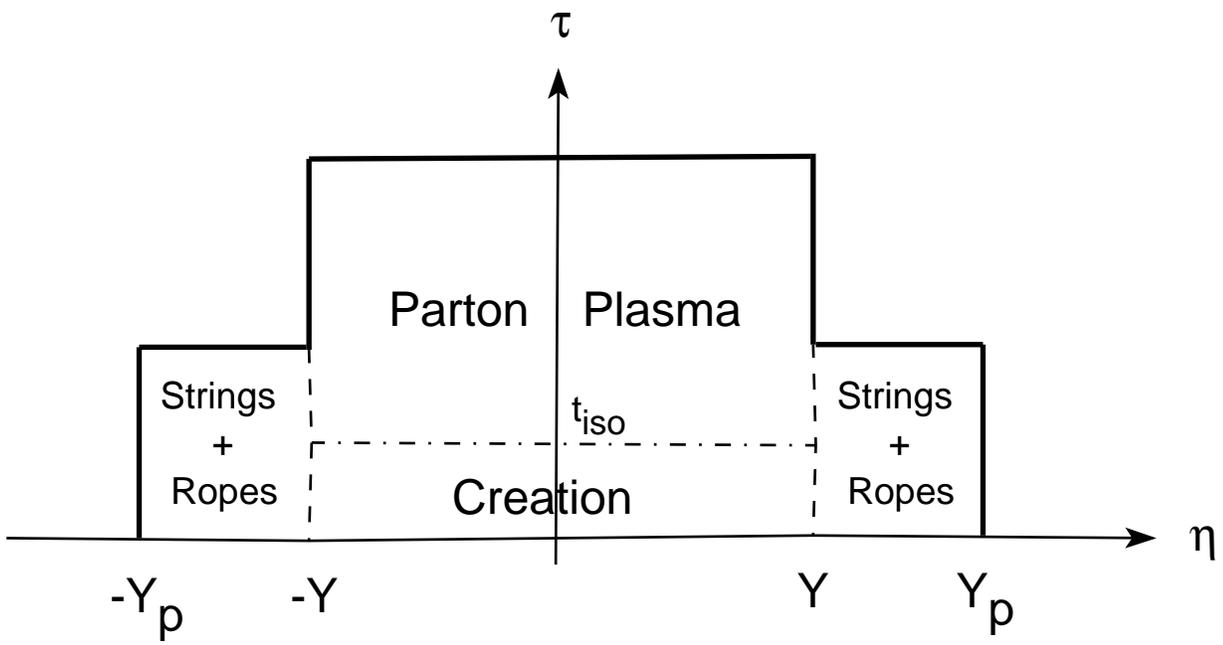
\end{document}